\documentclass[aps,preprint,amsmath,amssymb]{revtex4}
\usepackage{graphicx,color}
\usepackage{epsfig}
\def\be{\begin{eqnarray}}
\def\ed{\end{eqnarray}}
\def\non{\nonumber}
\def\ga{\gamma}

\begin{document}

\title{\Large \bf  Colored bosons on top FBA and angular cross section for $t \bar t$ production }

\date{\today}

\author{ \bf  R.~Benbrik $^{1,2}$\footnote{benbrik@ifca.unican.es}, Chuan-Hung Chen$^{5}$\footnote{physchen@mail.ncku.edu.tw} and M. EL Kacimi$^{3,4}$\footnote{elkacimi@uca.ma}
}

\affiliation{ $^{1}$Instituto de F\'isica de Cantabria (CSIC-UC), Santander, Spain\\
$^{2}$Facult\'e Polydisciplinaire de Safi, Sidi Bouzid B.P 4162, 46000 Safi,
Morocco\\
$^{3}$LPHEA, FSSM, Cadi Ayyad University, B.P. 2390, Marrakesh,
Morocco\\
$^{4}$LAPP, Laboratoire d'Annecy-Le-Vieux de Physique des Particules,
Annecy, France\\
$^{5}$Department of Physics, National Cheng-Kung
University, Tainan 701, Taiwan 
 }

\begin{abstract}

With full data set that corresponds  to an integrated luminosity of
9.4 fb$^{-1}$, CDF has updated the top quark forward-backward
asymmetry (FBA) as functions of rapidity difference $|\Delta y|$ and
$t\bar t$ invariant mass $M_{t\bar t}$. Beside the sustained
inconsistency between experiments and standard model (SM) predictions
at large $|\Delta y|$ and $M_{t\bar t}$, an unexpected  large first
Legendre moment with $a_1= 0.39\pm 0.108 $ is found.  In order to
solve the large top FBA, we study the contributions of  color triplet
scalar and color octet vector boson. We find that  the top FBA at
$|\Delta y| >1$  ($M_{t\bar t} > 450$ GeV) in triplet and octet
model could be enhanced to be around 40\% (30\%) and 26\% (20\%), whereas the first
Legendre moment is $a^{\bf Di}_1= 0.38$ and $a^{\bf Axi}_1= 0.23$, respectively. 

\end{abstract}

\maketitle

It is believed that the standard model (SM) is just
an effective theory of a fundamental one yet to be discovered, even it is consistent with most experimental data. 
 For pursuing more basic elements which are made of our universe, the search of new physics beyond SM is inevitable. If such new physics exists,
it can be probed either directly at collider or indirectly through
precise measurements. The recent measurements at
 Tevatron on the  forward-backward asymmetry (FBA)  in
 the top-quark pair production at $\sqrt{s}=1.96$ TeV may provide the clue
for the existence of  new physics. The FBA of top pair system is defined by 
  \be
  A_{FB}= \frac{N(\Delta y >0) - N(\Delta y <0)}{N(\Delta y >0) + N(\Delta y <0)}\, ,
  \ed
where $\Delta y=y_t - y_{\bar t}$, $y_{t(\bar t)}$ is the rapidity of
top (anti-top) quark and $N$ is the number of events for $\Delta y \gtrless 0$.

With full Tevatron data set which corresponds to an integrated luminosity of 9.4 fb$^{-1}$, CDF Collaboration measures the top-quark FBA  with lepton+jets topology at parton level to be \cite{Aaltonen:2012it}
 \be
 A_{FB}= 0.164 \pm 0.047 \,. \label{eq:CDF}
 \ed
The result  is roughly in agreement with early CDF and D0 data
\cite{D0_PRL100,CDF_PRL101}.  CDF also reports  the linear mass
($A_{FB}(M_{t \bar t} )$) and rapidity ($A_{FB}(|\Delta y|)$ )
dependences and the associated slope as  $(15.2\pm 5.0)\times 10^{-4}$
and $(28.6\pm 8.5)\times 10^{-2}$, respectively.  The former is
1.3$\sigma$ deviation from the SM prediction and the latter is
2.1$\sigma$. Additionally, CDF also analyzes the angular differential
cross section for $t\bar t$ production and observes an unexpected
large first Legendre moment, where 
it is associated with top FBA \cite{CDF-note-10974}. 

In the SM, since the top-quark pair production is
dominated by the strong interaction QCD contribution, due to
C-parity invariance, a vanishing FBA at the leading order (LO) is
expected. However, a nonvanishing FBA can be induced at the
next-to-leading order (NLO) \cite{AKR} and beyond \cite{Ahrens:2011uf,Hollik:2011ps,Kuhn:2011ri,Manohar:2012rs,LeiWang,Bernreuther:2012sx,Brodsky:2012ik}. The SM predictions have been improved and the resultant range could be from around 6\% to  around 10\% \cite{Brodsky:2012ik,Aguilar-Saavedra:2013rza}, 
however, by comparing with the central value of Eq.~(\ref{eq:CDF}),  the inconsistency between SM and data does not disappear.  Although the anomaly of FBA has derived many studies of  new physics in the literature \cite{Arhrib:2009hu,Frampton:2009rk,Antunano:2007da,NP1,NP2,NP3,NP4,NP5,NP6,Berger:2011ua,Jung:2009pi,Biswal:2012mr}, based on the new measurements and updated data, we investigate the issue by introducing new u-channel and s-channel effects. For illustration, we study  the color triplet \cite{Arhrib:2009hu} and color octet 
models \cite{Frampton:2009rk,Antunano:2007da,Choudhury:2007ux}, where the former is a representative of u-channel and the latter is a s-channel.

In order to study the angular cross section for $t\bar t$ production  and the unexpected large top FBA as functions of $|\Delta y|$ and $M_{t\bar t}$, here we consider two extensions of the SM. One is to introduce a color triplet scalar, called diquark, to the SM. Although there are many possible representations of diquark in SM gauge symmetry, for simplicity, we only focus on  the representation $(3, 1, -3/4)$. The other is color octet  gauge boson which could be arisen from an extended $SU(3)_R\times SU(3)_{L}$ gauge symmetry, called chiral color gauge model   \cite{chiral,nonuni}. In such model, the SM QCD could be taken as an unbroken symmetry of the extended one. Since the couplings of the new color gauge boson to quarks have the axial vector currents, hereafter, we call the massive color octet gauge boson as axigluon. 

 Now we briefly introduce the relevant pieces for the new models. Firstly, we discuss the color triplet model.  The SM gauge invariant interactions of color triplet diquark
with quarks are written by
 \be
 -{\cal L}_{\bf 3}&=&f^{\bf 3}_{ij} \bar u_{i\alpha} P_L u^c_{j\beta}
 \bar K^{\alpha \beta}_{\ga} H^{\ga\dagger}_{\bf 3} + h.c. \,, 
 \label{eq:color3}
 \ed
where the indices $i$ and $j$ are the quark flavors, $f^{\bf 3}_{ij}=-f^{\bf 3}_{ji}$,
$(\alpha,\beta,\gamma)$ stand for the color indices and $P_{L(R)}=(1\mp \gamma_5)/2$ is the helicity projection.
The antisymmetric tensors in color space are defined as $\bar K^{\alpha \beta}_{\gamma} = (K^\gamma_{\alpha\beta})^\dagger$ and the $K$s are given by \cite{Han:2009ya}
 \be
 K_1=\frac{1}{\sqrt{2}} 
  \left( \begin{tabular}{ccc}
  0 & 0 & 0 \\
  0 & 0 & 1 \\
  0 & -1 & 0
\end{tabular} \right)\,, \ \ \  K_2=\frac{1}{\sqrt{2}} 
  \left( \begin{tabular}{ccc}
  0 & 0 & -1 \\
  0 & 0 & 0 \\
  1 & 0 & 0
\end{tabular} \right)\,, \ \ \  K_3=\frac{1}{\sqrt{2}} 
  \left( \begin{tabular}{ccc}
  0 & 1 & 0 \\
  -1 & 0 & 0 \\
  0 & 0 & 0
\end{tabular} \right)\,,
 \ed
where the antisymmetric tensors satisfy  $Tr(K^a \bar K_{b} )= \delta^a_b$ and $K^a_{\alpha\beta} \bar K^{\rho \sigma}_{a} = \frac{1}{2} ( \delta^\sigma_\alpha \delta^{\rho}_{\beta} - \delta^{\rho}_{\alpha} \delta^{\sigma}_{\beta} )$. 
As a result,  the process $u\bar u \to t\bar t$ could proceed through the following interactions
 \be
- {\cal L}_{\bf 3}&=& 2f^{\bf 3}_{ut} \bar u_\alpha P_L t^c_\beta
 \bar K^{\alpha \beta}_{\ga} H^{\ga\dagger}_{\bf 3} + h.c.
 \label{eq:ut_int}
 \ed

For color octet gauge boson of $SU(3)_{R}\times SU(3)_{L}$, we write the interactions of the axigluon with quarks as
  \be
{\cal L}_{A} &=& g_V \bar q \ga_\mu {\bf Z_1}T^b q G^{b\mu}_{A} + g_A \bar q \ga_\mu \ga_5
{\bf Z_2} T^b q G^{b\mu}_{A}\,, 
  \ed
where the flavor and color indices are suppressed, $q^T=(u, c, t)$, $g_{V, A}$ are the gauge couplings of axigluon to the first two generation quarks, $T^b$ are the Gell-Mann matrices which are normalized by $Tr(T^b T^c)= \delta^{ac}/2$ and $\bf Z_i$ is a $3\times 3$ diagonalized matrix with diag($Z_i$)=(1, 1, $\zeta_i$). Here $\zeta_{1(2)} =g^t_V/g_V(g^t_A/g_A)$. $g^t_{V(A)}$ denotes the gauge coupling of the third-generation quark and its value depends on a specific model, e.g. $\zeta_{1(2)}=1(-1)$ in Ref.~\cite{Frampton:2009rk}. Although flavor changing neutral currents  (FCNCs) at tree level could be induced by $g^t_{V(A)} \neq g_{V(A)}$ and have an interesting influence on low energy physics \cite{Chen:2010wv}, since the small effects do not have a significant contributions to the studying phenomena, we will not further discuss the FCNC effects. 
In our numerical analysis, the parametrisation of free parameters is the same as that in Ref.~\cite{Frampton:2009rk}. 
Therefore, the gauge coupling of SM QCD is given by $g_s= g \sin\phi \cos\phi$, $\sin\phi= g_R/g$, $\cos\phi=g_L/g$, $g=\sqrt{g^2_R + g^2_L}$ is the combination of new gauge couplings, $g^t_V=g_V=-\frac{1}{2}g\cos2\phi$ and $g^t_A= -g_A=g/2$. If we take the value of $g_s$ as input, the mixing angle $\phi$ and  mass of axigluon are the main free parameters. 

Before presenting the scattering amplitude squares for $q\bar q\to t\bar t$ which are mediated by
gluon and the colored bosons, let us first define a convenient notation
for the following calculations.  The momenta of the incoming
quark and anti-quark, outgoing top and outgoing anti-top
quarks are denoted by $p_q$, $p_{\bar{q}}$ (q=u,d), $p_t$ and $p_{\bar{t}}$
respectively such that $p_q+p_{\bar{q}}=p_t+p_{\bar{t}}$. The
momentum can be written as:
\begin{eqnarray}
&&p_{q,\bar{q}}= \frac{\sqrt{\hat s} }{2}(1,0,0,\pm 1)\nonumber\\
&&p_{t,\bar{t}}= \frac{\sqrt{\hat s} }{2}(1,\pm\beta \sin\theta,0,\pm
\beta\cos\theta)
\label{eq:mom}
\end{eqnarray}
where $\beta^2 = 1-4m_t^2/\hat s$ and $\theta$ is
the scattering angle in the center-of-mass frame of the $t\bar t$.
Neglecting the light quark masses of the incoming partons,
the Mandelstam variables are defined as follows:
\begin{eqnarray}
\hat s &=&  (p_q+p_{\bar{q}})^2 = (p_t+p_{\bar{t}})^2\,,  \nonumber \\
\hat t &=& (p_q-p_t)^2 = (p_{\bar{q}}- p_{\bar{t}})^2
= m_t^2 - \frac{\hat s}{2} \left( 1 - \beta \cos\theta
                                              \right )\,,
\nonumber \\
\hat u &=& (p_q-p_{\bar{t}})^2 = (p_{\bar{q}}-p_t)^2
=m_t^2 - \frac{\hat s}{2} \left( 1 + \beta \cos\theta
                                              \right )\,.
\label{eq:mandel}
\end{eqnarray}
Accordingly,  the averaged amplitude square for QCD gluon and color triplet can be derived by \cite{Arhrib:2009hu}
 \be
\overline{\sum  \left| {\cal M}_{SM+H_3} \right |^2 } &=& \frac{4\pi^2
\alpha_s^2}{N^2_c} (N^2_c-1) \left(1+\beta^2
\cos^2\theta + \frac{4m^2_t}{\hat{s}} \right) \non \\
&+& \frac{\pi \alpha_s}{N^2_c} \left( \frac{N^2_c-1}{2}\right)
\frac{f^{\prime^ 2}_{ut}\hat{s} }{\hat{u}-m^2_{H_3}} \left((1+\beta\cos\theta)^2
+ \frac{4m^2_t}{\hat{s}} \right) \non \\
&+& \frac{1}{8N^2_c}N_{c} (N_c -1 ) \frac{f^{\prime^ 4}_{ut}
\hat{s}^2}{(\hat{u} -m^2_{H_3})^2}  (1+\beta\cos\theta)^2 \,,
 \label{eq:amp2_3}
 \ed
where we have already summed over final state color and averaged
over the initial spin and color, $m_{H_3}$ is the mass of diquark,
$f'_{ut}=2f_{ut}$ and $N_c=3$. 
We note
that since the propagator in the $u$-channel diagram depends on the
scattering angle $\theta$,
the FBA may arise not only from $\cos\theta$ terms in
Eq.~(\ref{eq:amp2_3}), but also from the constant terms and the
$\cos^2\theta$ terms. Although the interference  between SM
and diquark is negative when
$\hat{u}-m^2_{H_3}<0$, however the pure diquark contribution is positive and dominates in the considered $M_{t\bar{t}}$ range.  
 

The averaged amplitude square for QCD gluon and  axigluon contributions can be obtained as
 \be
\overline{ \sum  \left| {\cal M}_{SM+G_A} \right |^2 } & =& \frac{N^2_c-1}{4N^2_c} \left\{
16\pi^2 \alpha^2_s \left( 1 + \beta^2 \cos\theta^2 +  \frac{4m^2_t}{\hat s} \right)  \right. \non \\
&+&   \frac{8\pi \alpha_s \hat s (\hat s - m^2_A)}{(\hat s - m^2_A)^2+m^2_A \Gamma^2_A} 
 \left[ g_V g^t_V  \left( 1 + \beta^2 \cos^2\theta + \frac{4 m^2_t}{\hat s} \right)  + 2 g_V g^t_A \beta \cos\theta \right] \non \\
 &+& \frac{\hat s^2}{ (\hat s - m^2_A)^2+m^2_A \Gamma^2_A} \left[ \left( g^2_V + g^2_A\right) 
 \left( (g^t_V)^2 \left( 1+ \beta^2 \cos^2\theta + \frac{4m^2_t }{\hat s} \right) \right. \right. \non\\
 &+& \left.  \left. \left.  (g^{t}_A)^2 \left( 1+ \beta^2 \cos^2\theta - \frac{4m^2_t }{\hat s} \right) \right)  
 +8 g_V g^t_V g_A g^t_A \beta \cos\theta \right] \right\}.\label{eq:amp2_8}
 \ed 
Clearly, the FBA is only from 
the linear $\cos\theta$ terms  and  associated with axial-vector
coupling. Unlike diquark model, 
with $g_Vg^t_A,\, g_V g^t_V g_A g^t_A <0$, 
the interference between SM and axigluon is
positive when $\hat{s}-m^2_A<0$,  whereas axigluon contribution  is negative.  Therefore, both
contributions to the FBA are in competition.

Beside the experimental limit on the mass of new particles, the direct strict constraint on the free parameters is from the $t\bar t$ production cross section. Based on 8.8 fb$^{-1}$ of Tevatron data, the recent combination of CDF and D0 results for $t\bar t$ cross section is 
\cite{Deliot:2013gla}
\begin{equation}
  \sigma (p\bar p \to t \bar t)^{\rm exp} = 7.65\pm 0.20 \pm 0.36
  \ \ {\rm pb}\,.
\label{eq:data}
\end{equation}
In order to study new physics on $t \bar t$ production cross section and top FBA,
we implement the diquark 
and axigluon matrix elements, shown in Eqs. (\ref{eq:amp2_3}) and (\ref{eq:amp2_8}), in 
Pythia8175 \cite{Pythia8:ref} as semi-internal $2\to 2$ processes. Thus, $t\bar{t}$ pair production in
 the two models could be dealt with as a normal internal Pythia process. 
For estimating the SM NLO contributions,  we use
POWHEG-BOX-1.0 \cite{Powheg:ref}. 



All estimations in our analysis are performed at parton level
and we
do not take into account the effects from parton showering,
hadronizations and detector conditions.
The taken inputs are the experimental value
given by Eq.~(\ref{eq:data}) within $2\sigma$ errors, $m_{t} = 172.5$
GeV and $\alpha_s = 0.1095$
at $m_Z$. 
The renormalization and factorisation scales are fixed 
at $\mu_R=\mu_F=m_t$. The event samples are generated
using CTEQ6L1~\cite{cteq6:ref} set parton distribution functions   for
diquark and axigluon models, 
whereas CTEQ6M~\cite{cteq6:ref} is used for SM NLO.

\begin{figure}[hptb]
\includegraphics[scale=0.55] {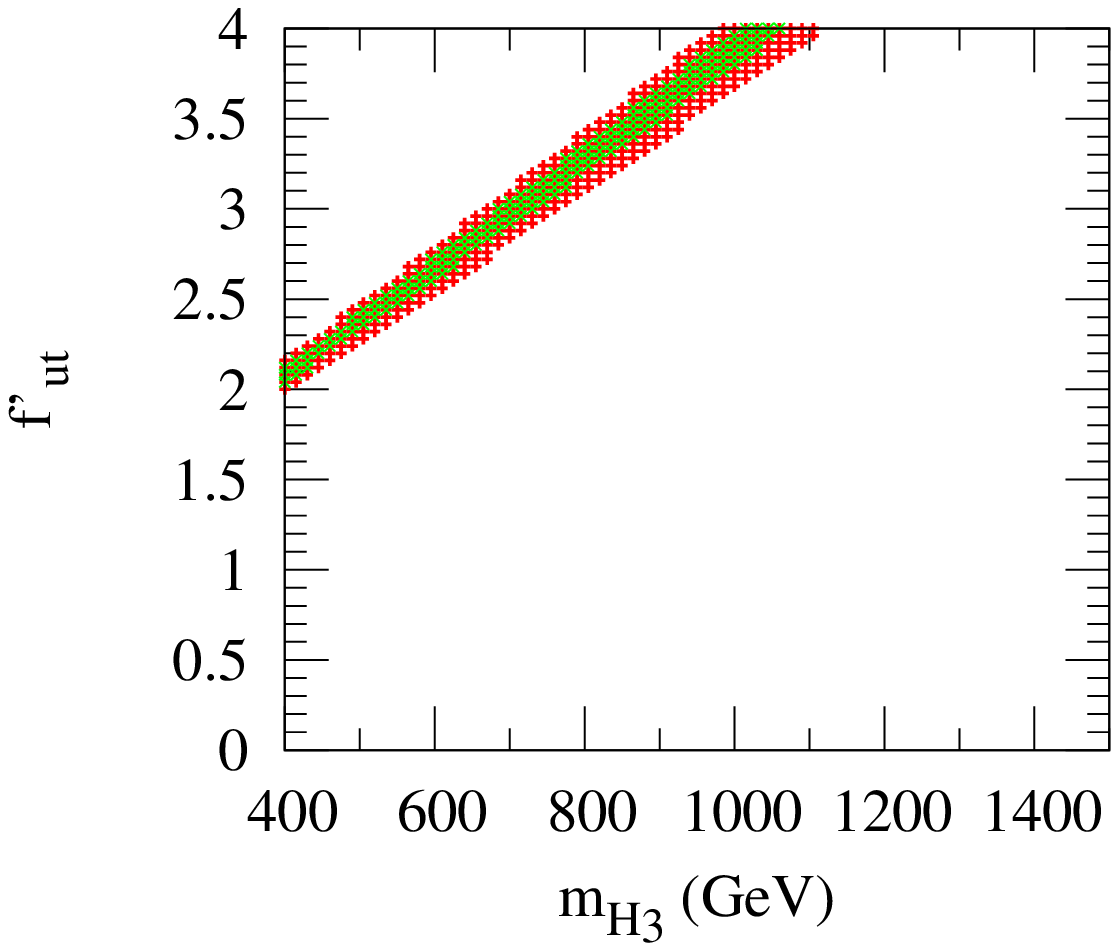}
 \includegraphics[scale=0.55] {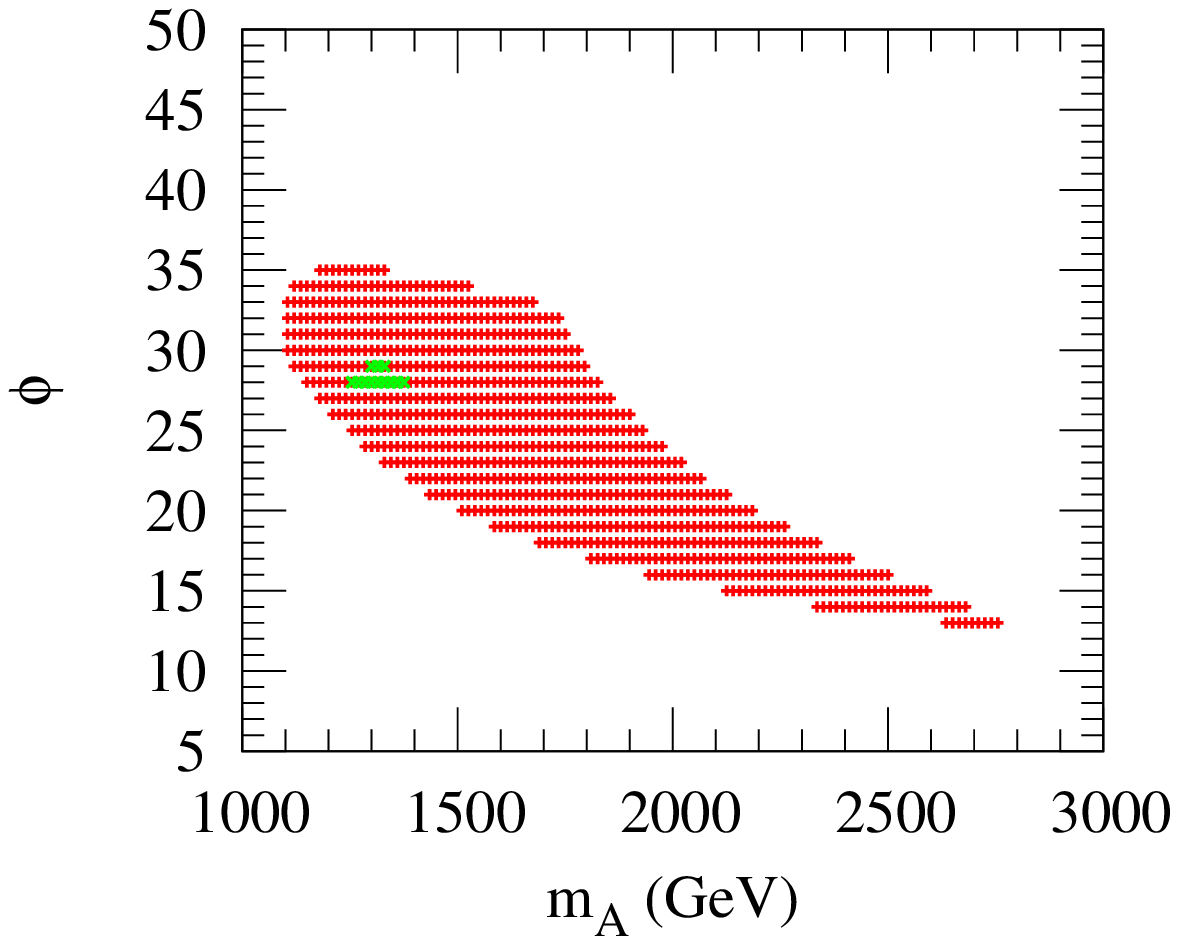}
\caption{Scatter plots of $t\bar t$ production cross section for color
  triplet diqaurk (left panel) and axigluon (right panel) within
  1$\sigma$ (green) and 2$\sigma$(red) of
  $\sigma^{exp}_{t\bar{t}}$ and $A_{FB}$, respectively.}
\label{fig:fig1}
\end{figure}

 Before we discuss the contributions of new physics to top FBA, we first study the allowed parameter space, where the main constraint is from the $t\bar t$ production cross section given  in Eq.~(\ref{eq:data}). For color triplet, since the free parameters are $f_{ut}$ and $m_{H_3}$, we display the scatters of $\sigma( p \bar p \to t \bar t)$ with respect to $f'_{ut}=2f_{ut}$ and $m_{H_3}$ in left panel of Fig.~\ref{fig:fig1}. Similarly, for axigluon model, the scatters of $\sigma( p \bar p \to t \bar t)$ with respect to  mixing angle $\phi$ and $m_{A}$ are shown in right panel of Fig.~\ref{fig:fig1}.  

\begin{figure}[hptb]
\includegraphics[scale=0.40] {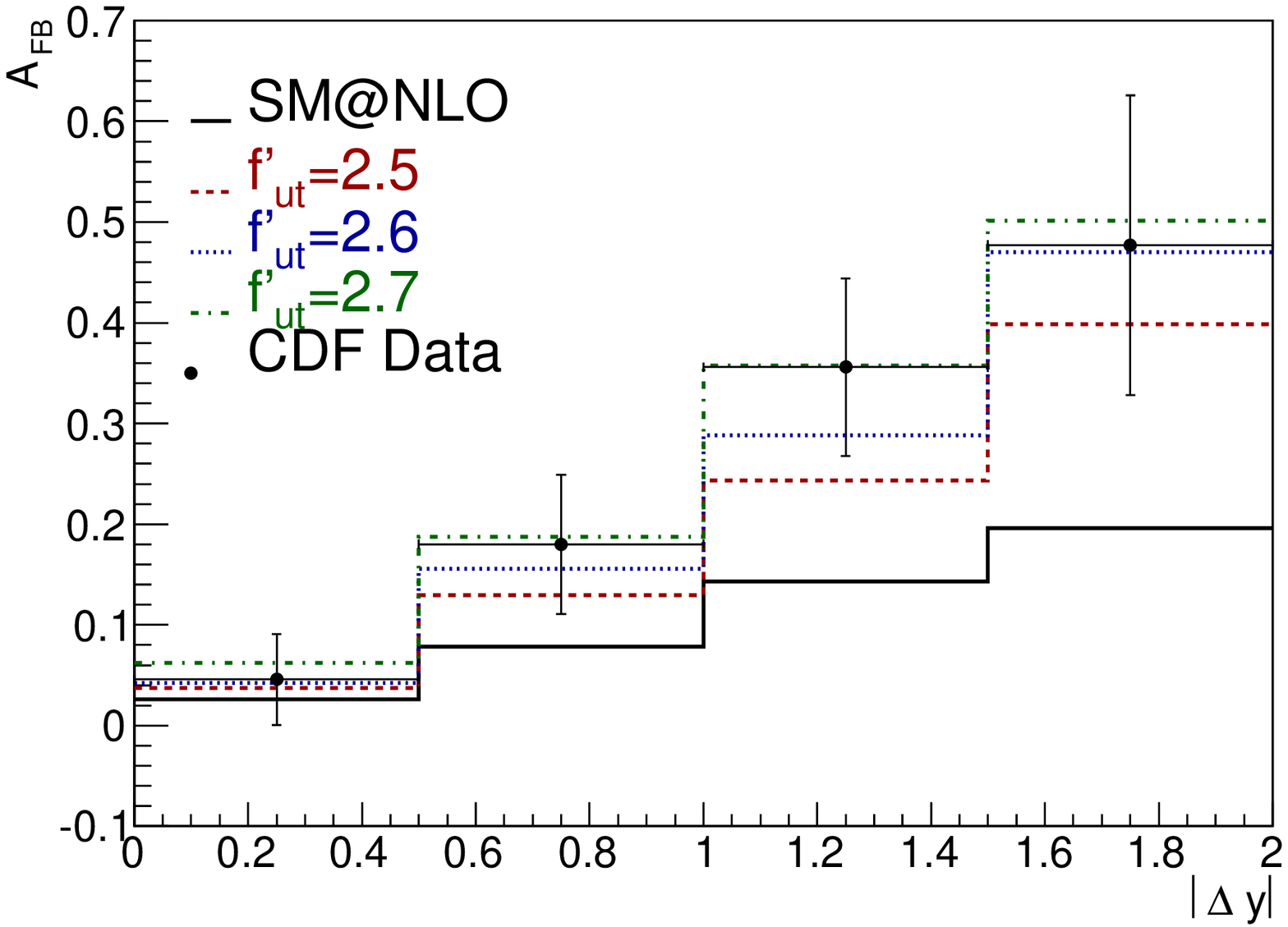}
\includegraphics[scale=0.40] {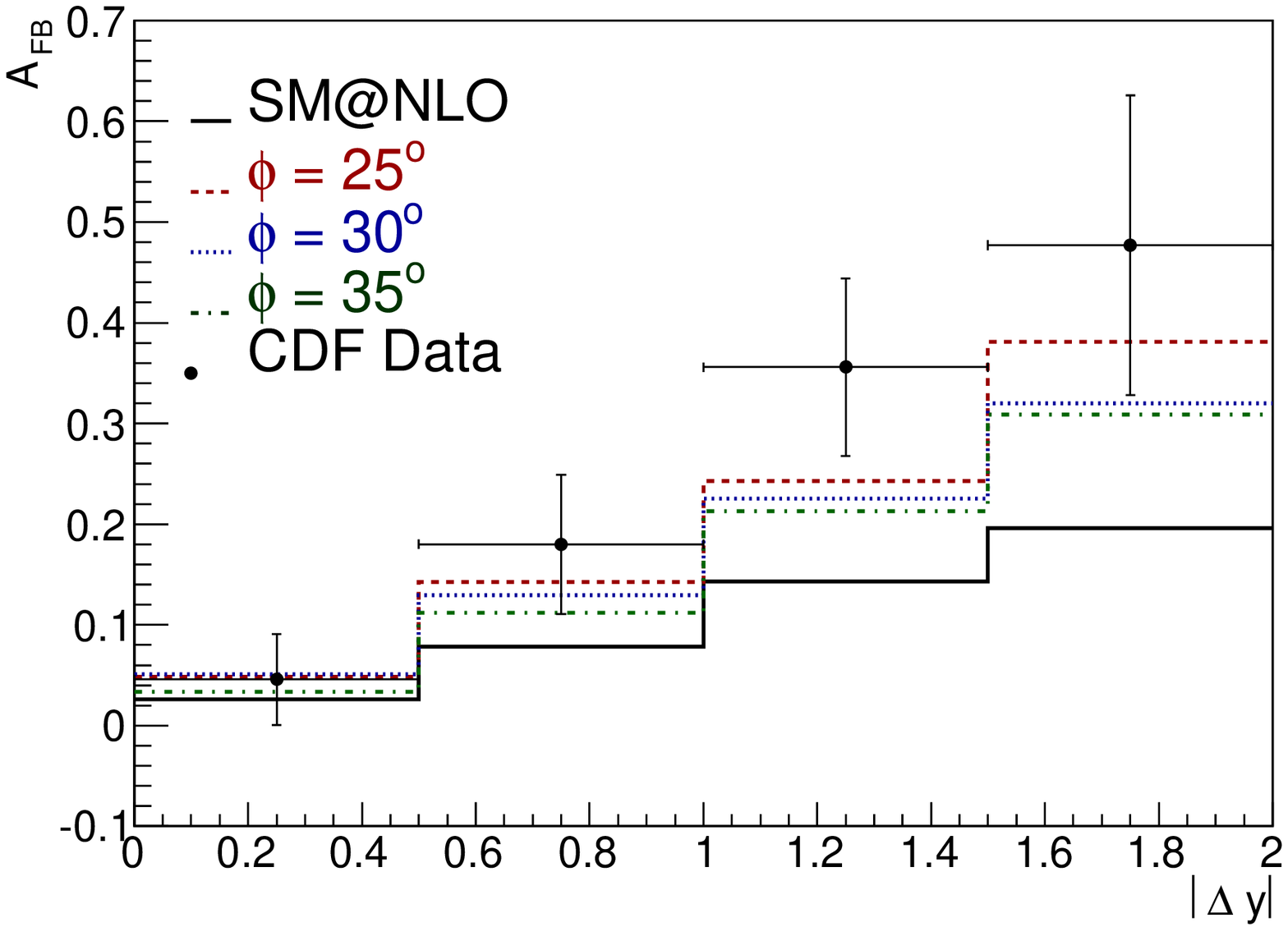}
\caption{Top FBA as a function of $|\Delta y|$, where the dashed, dotted and dash-dotted lines stand for $f'_{ut}=(2.5, 2.6, 2.7)$ in diquark model and $\phi=(25^0, 30^0 ,35^0)$ in axigluon model, respectively.   The solid line is the SM NLO.}
\label{fig:fig2}
\end{figure}

With the allowed parameters shown in
  Fig.~\ref{fig:fig1}, we now discuss the influence of color triplet
  and octet bosons on the FBA. In order to display the dependence of
  free  parameters, we will fix the masses of new bosons and chosen
  the allowed values for the couplings. Consequently, with
  $m_{H_3}=665$ GeV,  the top FBA as a function $|\Delta y|$ for diquark
  is shown in left panel of Fig.~\ref{fig:fig2}. The dashed, dotted and
  dott-dashed lines stand for $f'_{ut}=(2.5,
  2.6,2.7)$, where the corresponding $t\bar t$ production rates are
  $\sigma(p \bar p \to t \bar t)=(7.29,7.70,8.19)$ pb. 
 For axigluon model, we present the results in right panel of Fig.~\ref{fig:fig2}. For
escaping the limit from the search of  new resonance, we take
$m_{A}=1525$ GeV.  The dashed, dotted and dash-dotted  lines
denote  $\phi=(25^{0}, 30^{0}, 35^{0})$, where
the corresponding $t\bar t$ cross sections are $\sigma(p \bar p \to t
\bar t) = (7.35,7.72,7.94)$ pb.
 For comparisons, we also show the SM NLO prediction in the plots by solid line. By the two plots, we see clearly both models could enhance the top FBA and match the CDF data. 

\begin{figure}[hptb]
\includegraphics[scale=0.40] {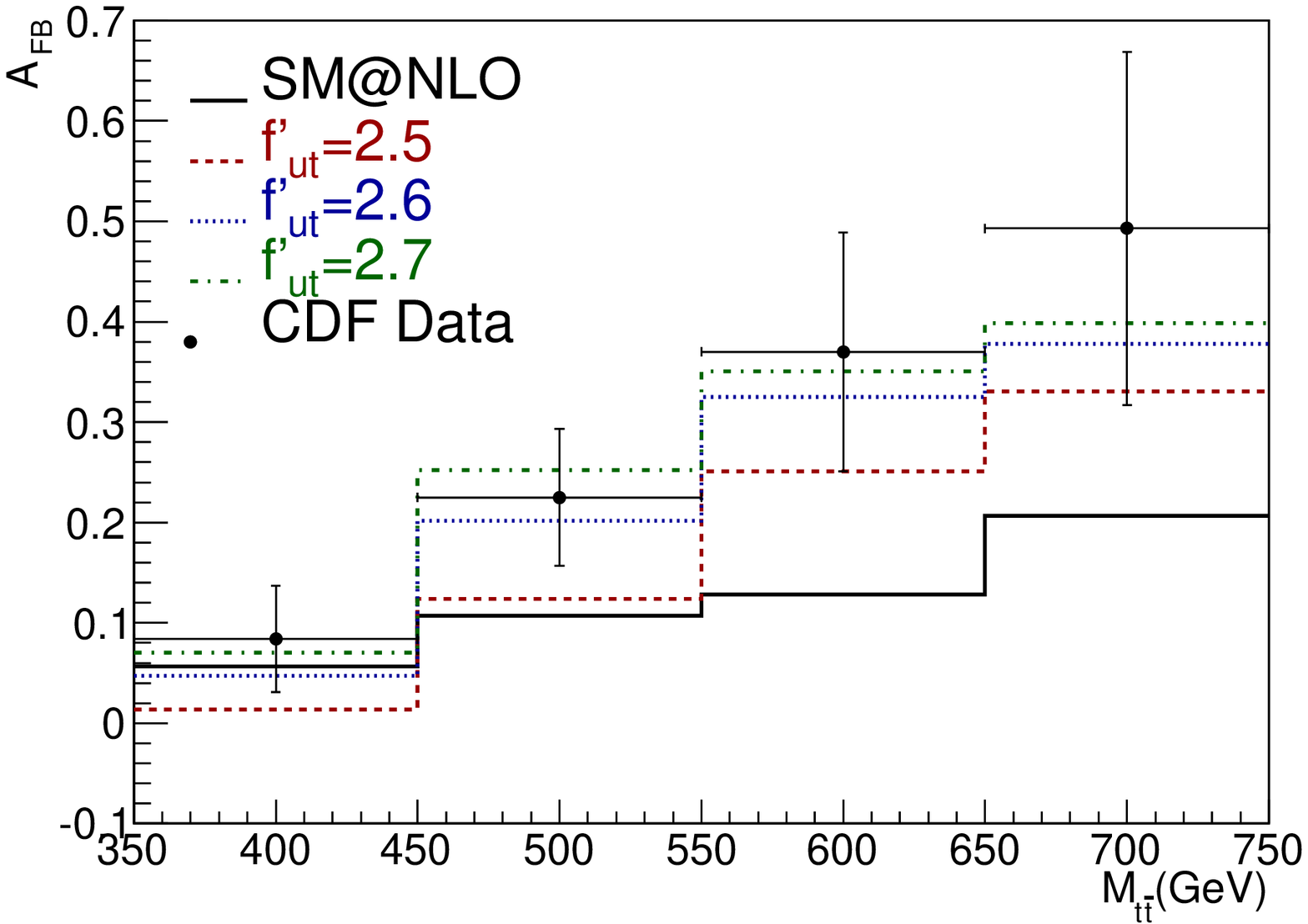}
\includegraphics[scale=0.40] {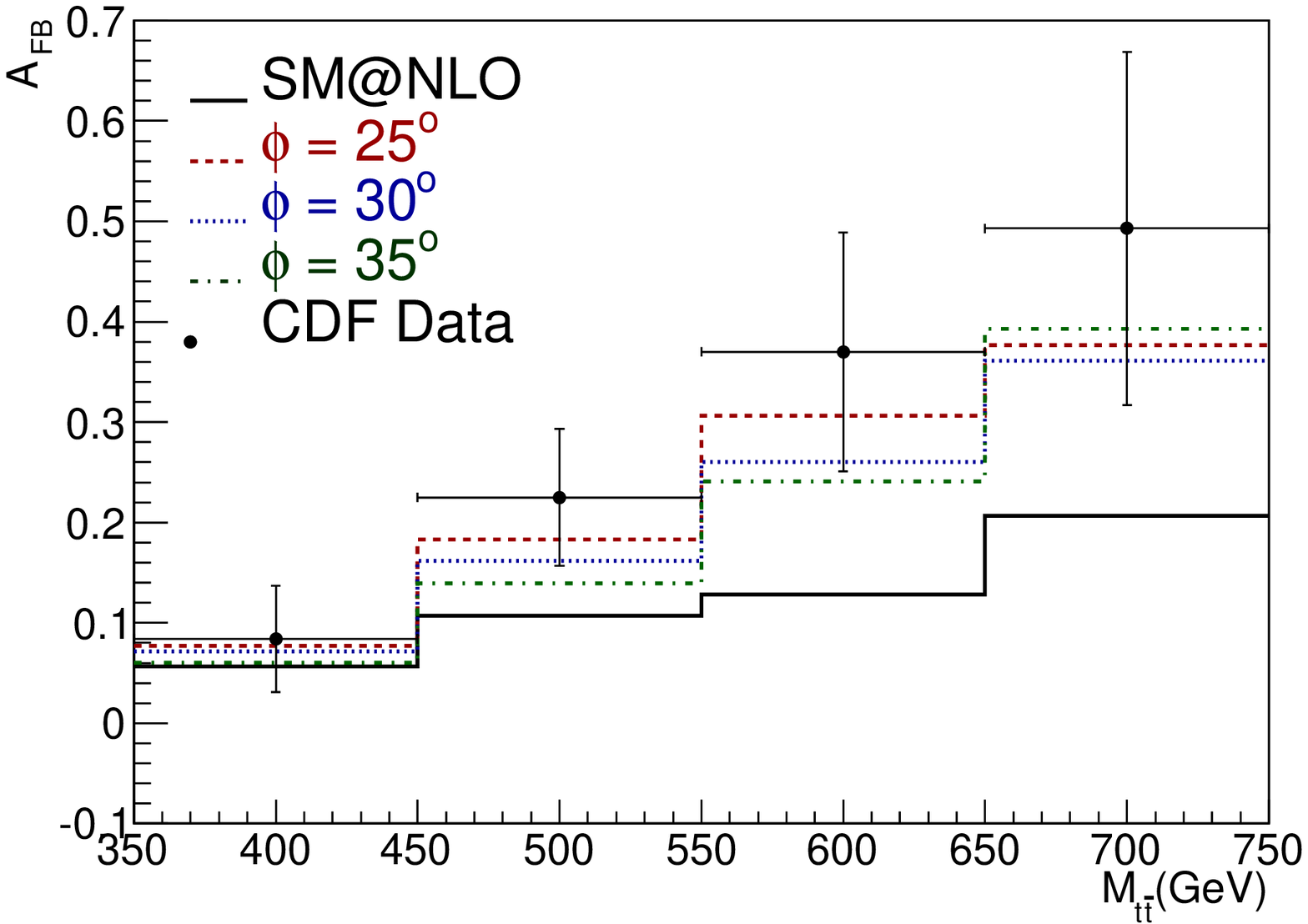}
\caption{  The legend is the same as Fig.~\ref{fig:fig2} but for function of $M_{t\bar t}$. }
\label{fig:fig3}
\end{figure}

\begin{figure}[hptb]
\includegraphics[scale=0.55] {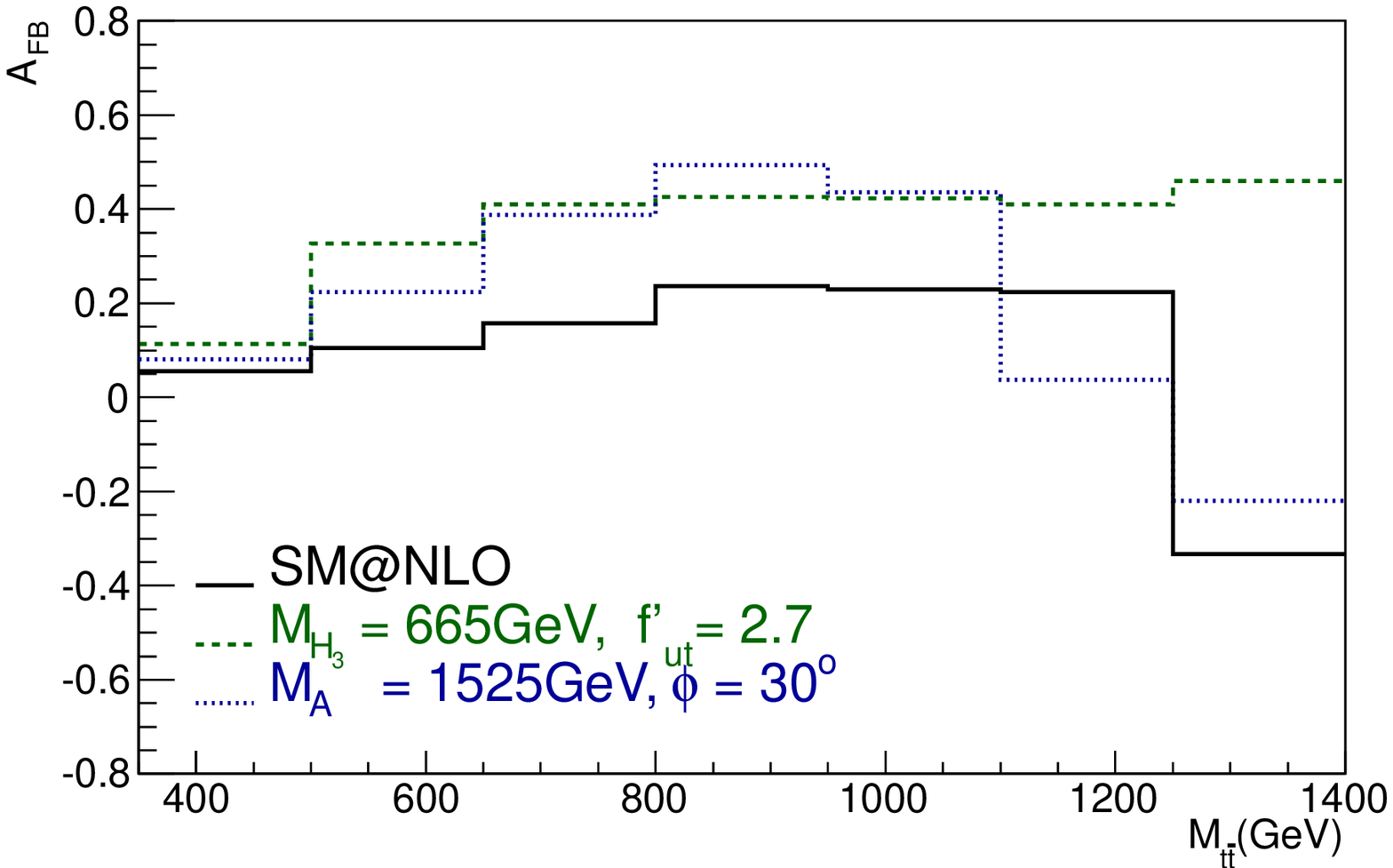} 
 \caption{ Differential asymmetry $A_{FB}$ as a function of
   $M_{t\bar{t}}$ integrated over 7 bins chosen in  $[350,1400]$ GeV.
 }
\label{fig:afbvsmttp7}
\end{figure}

Using the same taken values of parameters for $|\Delta y|$ dependence, we present the FBA as a function of $M_{t\bar t}$ in Fig.~\ref{fig:fig3}.
The FBA in both models is enhanced at $M_{t\bar t}>450$ GeV and fits well to current CDF data within $1\sigma$ errors. 
 For further comparing our results with data and SM NLO, we show the values of FBA with the chosen ranges of $|\Delta y|$ and $M_{t\bar t}$ in Tables \ref{tab:table_dy} and \ref{tab:table_Mtt}.
\begin{table}[hptb]
\caption{  The asymmetry as a function of $|\Delta y|$ in color
  triplet and octet models. The data are quoted from
  \cite{Aaltonen:2012it}.  $m_{H_3}=665$ GeV, $f'_{ut}=2.7$,
  $m_A=1525$ GeV, $\phi=30^{0}$ are used for numerical calculations. 
 } \label{tab:table_dy}
\begin{ruledtabular}
\begin{tabular}{ccccc}
  $|\Delta y|$ & Data & SM@NLO & $H_{3}$ & $G_A$  \\ \hline
$0.0-0.5$ &   $0.048\pm 0.034 \pm 0.022$ &0.025  & $0.060  $ & $ 0.042$ \\
 $0.5-1.0$ & $0.180\pm 0.057 \pm 0.046$  &0.071  & $ 0.194$ &$ 0.126$ \\
 $1.0-1.5$ & $0.356 \pm 0.080 \pm 0.036$ &0.113& $ 0.352$ & $ 0.219$ \\
 $ \geq 1.5$ & $0.477\pm 0.132 \pm 0.074$ &0.171 & $0.562$ &  $0.370  $ \\ \hline
$<1.0 $ & $0.101\pm 0.040 \pm 0.029$ &0.042 & $ 0.118$ & $0.077 $ \\
 $ \geq 1.0$ & $0.392\pm 0.093 \pm 0.043$ &0.131 & $0.432 $ &  $0.266 $ \\ \hline 
 
\end{tabular}\end{ruledtabular}
\end{table}
\begin{table}[hptb]
\caption{ The legend is the same as Table \ref{tab:table_dy} but for $M_{t\bar t}$ dependence.  
} \label{tab:table_Mtt}
\begin{ruledtabular}
\begin{tabular}{ccccc}
  $M_{t\bar t}$ & Data & SM@NLO &  $H_{3}$ & $G_A$ \\ \hline
$<450$ &   $0.084\pm 0.046 \pm 0.026$ &0.048& $ 0.075 $ & $ 0.064$ \\
 $450-550$ & $0.255\pm 0.062 \pm 0.034$ &0.085& $ 0.245$ &$ 156$ \\
 $550-650$ & $0.370 \pm 0.084 \pm 0.087$ &0.115 & $0.358 $ & $ 0.263$\\
 $ \geq 650$ & $0.493\pm 0.158 \pm 0.110$ &0.170& $0.414 $ &  $0.398 $ \\ \hline
$<450 $ & $0.084\pm 0.046 \pm 0.030$ &0.048& $ 0.075$ & $0.064 $ \\
 $ \geq 450$ & $0.295\pm 0.058 \pm 0.033$&0.099 & $0.313 $ &  $ 0.205 $ \\ \hline 
 
\end{tabular}
\end{ruledtabular}
\end{table}
Beside  $M_{t\bar t} \leq 750$ GeV which is presented in CDF paper
\cite{Aaltonen:2012it}, we display the FBA up to 1400 GeV in
Fig.~\ref{fig:afbvsmttp7},
 where we have integrated the FBA 
over the width of 7 bins in [350, 1400] GeV. 
 It is found that when $M_{t\bar{t}}< 1100$ GeV, the two models induce a
positive asymmetry. When $M_{t\bar{t}}> 1100$ GeV, the asymmetry induced
by the diquark is positive and grows with $M_{t\bar{t}}$, whereas 
that induced by the axigluon is negative and falls with 
$M_{t\bar{t}}$ \cite{Frampton:2009rk}. Thus we can use the different behavior at large
$M_{t\bar{t}}$  to rule out one of the two models.

\begin{figure}[hptb]
\includegraphics[scale=0.4] {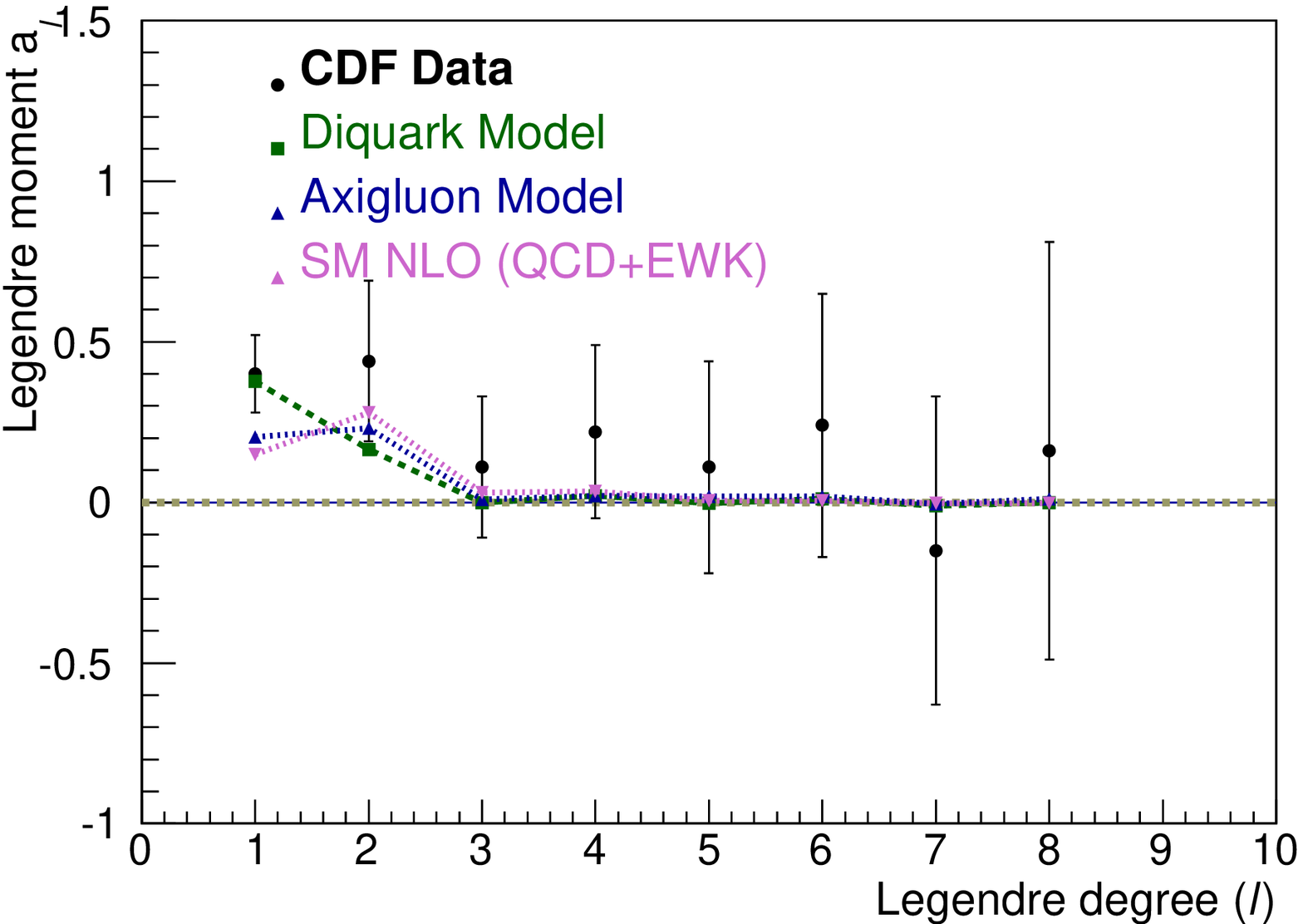}  
\includegraphics[scale=0.4] {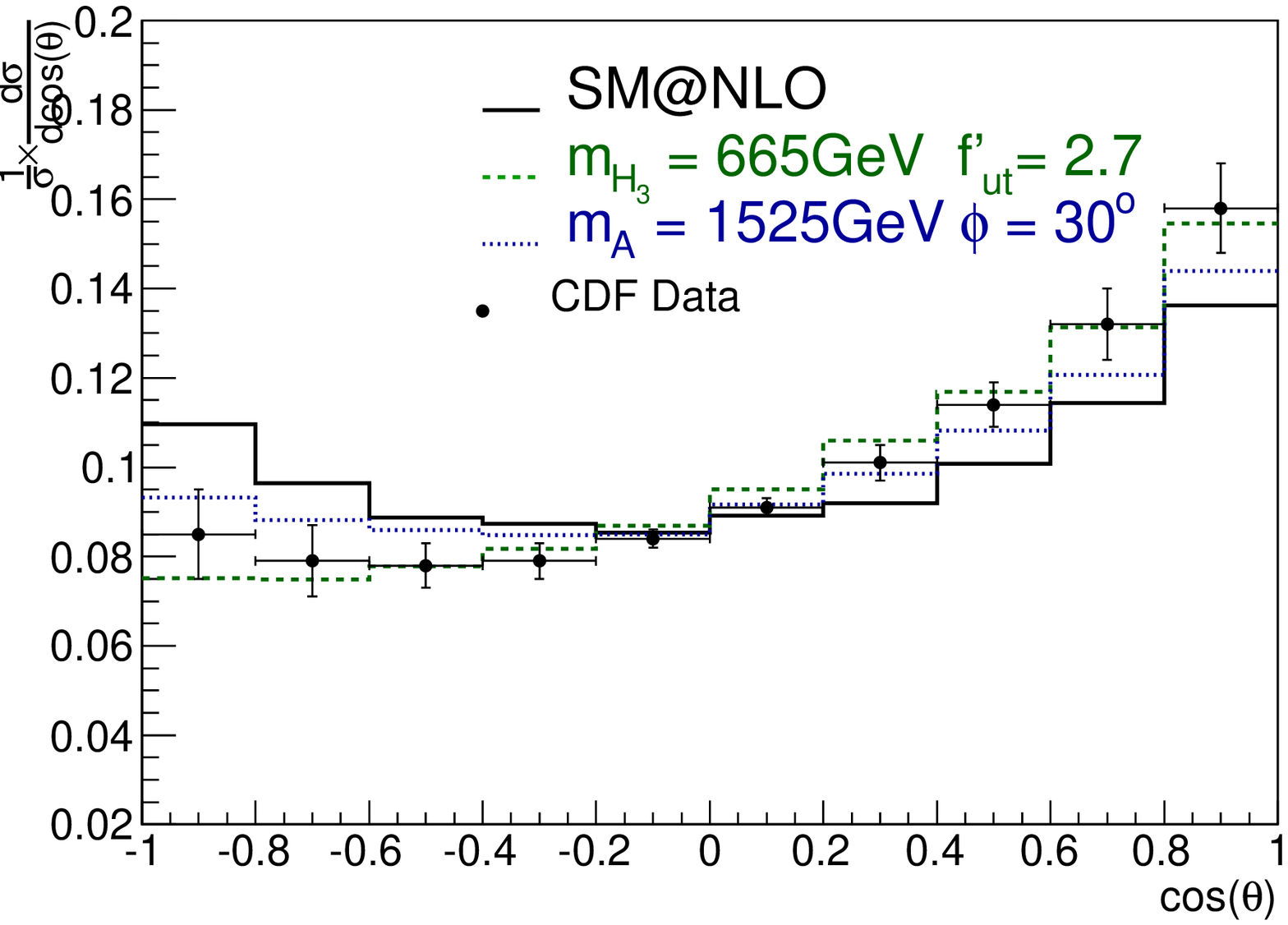}  
 \caption{ Left panel: Legendre moments estimated for diquark and
   axigluon models compared to SM NLO and CDF results. Right panel: normalised angular differential cross section for $p\bar p \to t \bar t$ as a function of $\cos\theta$, where the differential cross section has been integrated over 10 bins chosen in $[-1, 1]$.   }
\label{fig:fig_LM}
\end{figure}

CDF recently reports the measurements of the Legendre moments for
differential cross section of $t\bar t$ production with respect to the
scattering angle $\theta$ of top-quark in the $t\bar t$ center of mass. The moment is defined by 
 \be
 a'_{\ell} &=& \frac{2\ell +1}{2} \int^{1}_{-1} d\cos\theta
 \frac{d\sigma(\cos\theta)}{d\cos\theta} P_{\ell}(\cos\theta)\,,  \non \\
 &=& \frac{2\ell +1}{2}\sigma<P_{\ell}(\cos\theta)>
\label{eq:ael_0}
 \ed
with $\ell$ being the degree of Legendre polynomial. In order to compute the Legendre moments,  we normalize $a'_0$ to be unity. Thus, the Legendre moments  $a_\ell$ can be estimated
from a sample of $N_{ev}$ events  as
\be
a_{\ell} &=&(2\ell +1)< P_{\ell}(\cos\theta)> \non \\ 
&=&\frac{2\ell +1}{N_{ev}}\sum_{i=1}^{N_{ev}}  P_{\ell}(\cos\theta_i).
\ed 
It is found the first Legendre moment is $a_{1}=0.39 \pm 0.108$
\cite{CDF-note-10974} and the result is in disagreement with SM NLO of
$a^{\rm SM}_1=0.15^{+0.066}_{-0.033}$~\cite{Bernreuther:2012sx,CDF-note-10974}.
In our calculations, the first Legendre moment of diquark and axigluon model at the
preference point respectively is
\be
 a^{\bf Di}_1 &=& 0.38\,, \non \\
 a^{\bf Axi}_1 &=&0.23\,. 
 \ed
Clearly, diquark fits well CDF result. The first eight  Legendre moments $a_\ell$, $\ell$=1-8, are shown in the left panel of
Fig~\ref{fig:fig_LM}. For understanding the new physics contributions, we calculate the normalised differential cross section as a function of $\cos\theta$ and present the results in right panel of  Fig.~\ref{fig:fig_LM}, where the solid, dashed and dotted lines represent the SM NLO,  diquark with $m_{H_3}=665$ GeV and $f'_{ut}=2.7$ and axigluon with $m_A=1525$ GeV and $\phi=30^0$, respectively. 
In order to compare with CDF data, in the calculations we divide $[-1,1]$ into 10 bins and
integrate the angular differential cross section over the width of 10-bin
in $\cos\theta$. The right plot in Fig.~\ref{fig:fig_LM} shows that the diquark mode is in better
agreement with the data than axilguon model.  Note that, the
difference in the calculations of FBA between  $t\bar{t}$ frame and 
partonic center-of-mass frame is numerically negligible.

In summary, we have studied the top FBA in diquark and axigluon models, where the former is a representative of a u-channel and the latter is a s-channel. According to our analysis, both models could enhance the FBA and  fit well in $|\Delta y|$ and $M_{t\bar t}$ distributions. We also show that  the top FBA induced by s-channel will decrease from positive to negative at  $M_{t\bar t} > 1100$ GeV, while the u-channel is still growing slightly. We also give the first eight Legendre moments in diquark and axigluon models and find that diquark could explain the large $a_1$ obtained by CDF. \\

\noindent{\bf Acknowledgments}

We would like to thank Dr. Y.C. Chen of IOP Academia Sinica for useful
discussions. This work is supported by the National Science Council of
R.O.C. under Grant \#: NSC-100-2112-M-006-014-MY3 (CHC). 
The work of R.B was supported by the Spanish Consejo Superior de
Investigaciones Cientificas (CSIC) and by a part from National Center
for Theoretical Sciences (NCTS) where this work was initiated. The
work of M.E was supported by LIA
 { \it International Laboratory for Collider Physics} between IN2P3/CNRS, France, the Moroccan CNRST and KTH Sweden; LAPP, Universit\'e Savoie,
IN2P3/CNRS.  


\end{document}